\documentclass[aps, pre, amsmath, amssymb, showpacs,showkeys,floatfix,superscriptaddress,twocolumn]{revtex4}

\usepackage{natbib}
\usepackage{graphicx}
\usepackage{bm}
\usepackage{color}
\usepackage{epic, eepic}
\usepackage{subfigure}

\newcommand{\ddt}[1]    {\partial_t {#1} }
\newcommand{\ddx}[2]    {\partial_{#1} {#2} }
\newcommand{\ddxi}[1]   {\partial_i {#1} }
\newcommand{\ddxj}[1]   {\partial_j {#1} }
\newcommand{\ddxsq}[2]  {\partial_{#1}^2 {#2} }
\newcommand{\ddxjsq}[1] {\partial_j^2 {#1} }

\newcommand{\abs}[1] {\left| {#1} \right|}

\newcommand{\dddthat}[1]{\frac{d {#1}}{d \hat{t}}}   

\newcommand{\plav}[1]   {\langle {#1} \rangle_A}

\newcommand{\yav}[1]    {\langle {#1} \rangle_y}

\newcommand{\av}[1]     {\overline{#1}}
\newcommand{\fl}[1]     {#1'}
\newcommand{\sav}[1]      {\widetilde{#1}}
\newcommand{\sfl}[1]      {\fl{#1}}

\newcommand{\bA}[0]     {\mathcal{A}}
\newcommand{\bD}[0]     {\mathcal{D}}
\newcommand{\bP}[0]     {\mathcal{P}}
\newcommand{\bB}[0]     {\mathcal{B}}
\newcommand{\bR}[0]     {\mathcal{R}}

\newcommand{\ignore}[1]{}

\renewcommand{\Re}[0]     {\mathrm{Re}}
\newcommand{\Ra}[0]     {\mathrm{Ra}}
\newcommand{\Nu}[0]     {\mathrm{Nu}}
\renewcommand{\Pr}[0]     {\mathrm{Pr}}

\begin{document}

\title{Wind and boundary layers in Rayleigh-B\'{e}nard convection. \\
II. Boundary layer character and scaling}

\author{Maarten \surname{van Reeuwijk}}
\affiliation{Department of Civil and Environmental Engineering, Imperial College London, Imperial College Road, London, SW7 2AZ, UK}
\email{m.vanreeuwijk@imperial.ac.uk}

\author{Harm J.J. \surname{Jonker}}
\affiliation{Department of Multi-Scale Physics and
             J.M. Burgers Center for Fluid Dynamics,
             Delft University of Technology,
             Lorentzweg 1, 2628 CJ Delft, The Netherlands}

\author{Kemo \surname{Hanjali\'{c}}}
\affiliation{Department of Multi-Scale Physics and
             J.M. Burgers Center for Fluid Dynamics,
             Delft University of Technology,
             Lorentzweg 1, 2628 CJ Delft, The Netherlands}
\affiliation{Department of Mechanics and Aeronautics, University of Rome, ``La Sapienza'', Rome, Italy}

\date{December 27, 2007}



\begin{abstract}
The scaling of the kinematic boundary layer thickness $\lambda_u$ and the friction factor $C_f$ at the top- and bottom-wall of Rayleigh-B\'{e}nard convection is studied by Direct Numerical Simulation (DNS). By a detailed analysis of the friction factor, a new parameterisation for $C_f$ and $\lambda_u$ is proposed.
The simulations were made of an $L/H=4$ aspect-ratio domain with periodic lateral boundary conditions at $\Ra=\{10^5, 10^6, 10^7, 10^8\}$ and $\Pr=1$. 
A continuous spectrum, as well as significant forcing due to Reynolds stresses clearly indicates a turbulent character of the boundary layer, while viscous effects cannot be neglected judging from the scaling of classical integral boundary layer parameters with Reynolds number.
Using a conceptual wind model, we find that the friction factor $C_f$ should scale proportional to the thermal boundary layer thickness as $C_f \propto \lambda_\Theta$, while the kinetic boundary layer thickness $\lambda_u$ scales inversely proportional to the thermal boundary layer thickness and wind Reynolds number $\lambda_u \propto \lambda_\Theta^{-1} \Re^{-1}$. 
The predicted trends for $C_f$ and $\lambda_u$ are in agreement with DNS results.

\end{abstract}

\maketitle
\section{Introduction}

The structure of the boundary layer is of great importance for understanding the turbulent heat transfer characteristics of Rayleigh-B\'{e}nard convection. Inherently unstable due to buoyancy effects, the thermal boundary layer with thickness $\lambda_\Theta$ is in a dynamic equilibrium of heating (cooling)
by thermal diffusion and the detrainment (entrainment) of heat due to impinging and ejecting thermals at the bottom (top) plate. 
This process creates large temperature gradients across the boundary layer, thereby enhancing the heat transfer through the wall and thus the Nusselt number $\Nu$. Next to a thermal boundary layer, one can identify a kinematic boundary layer with thickness $\lambda_u$, associated with the velocity field. 
Depending on the Prandtl number $\Pr=\nu \kappa^{-1}$, which is the ratio between the kinematic viscosity $\nu$ and thermal diffusivity $\kappa$, the kinematic boundary layer can be nested inside the thermal boundary layer or vice versa, which influences the effectiveness of the heat transfer as a function of the Rayleigh number $\Ra$. The Rayleigh number $\Ra$ is defined as $\Ra = \beta g \Delta \Theta H^3 (\nu \kappa)^{-1}$, where $\beta$ is the thermal expansion coefficient, $g$ the gravitational constant, $\Delta \Theta$ the temperature difference between the top and bottom plate and $H$ the domain height.
The scaling of $\lambda_\Theta$ and $\lambda_u$ as a function of $\Ra$ and $\Pr$ are therefore of importance for proper prediction of the heat transfer.

In the theory of Grossmann and Lohse \citep{Grossmann2000}, the wind velocity $U$ and the boundary layer thicknesses $\lambda_u$ and $\lambda_\Theta$ are central parameters, which are used to estimate the dissipation rates of kinematic energy and temperature variance in the bulk and the boundary layers. 
In the theory, $\lambda_\Theta$ and $\lambda_u$ are defined as
\begin{gather}
  \label{eq:lambdat_GL}
  \lambda_\Theta \propto H /(2 \Nu), \\
  \label{eq:lambdau_GL}
  \lambda_u \propto H \Re^{-1/2}.
\end{gather}
While \eqref{eq:lambdat_GL} holds excellently, the correspondence of \eqref{eq:lambdau_GL} with experiments \citep{Xin1996, Xin1997} and simulations \citep{Kerr1996, Kerr2000} is less satisfactory.
Relation \eqref{eq:lambdau_GL} can be obtained by non-dimensionalising the steady laminar two-dimensional Prandtl boundary layer equations \citep{Schlichting2000, Grossmann2004}, by which \eqref{eq:lambdau_GL} follows immediately.
However, the measured $\Re$ dependence of $\lambda_u$ is much weaker than predicted by \eqref{eq:lambdau_GL} (see also Fig.\ \ref{fig:int_bl}).
It has been suggested that the difference is due to geometry effects \cite{Grossmann2003} (plate-filling vs.\ laterally restricted flow).

In this paper, we argue that the disparity in the expected and the observed scaling is because the top- and bottom boundary layers are not laminar, i.e. forcing due to Reynolds stresses cannot be neglected in the kinematic boundary layer.
Consequently, the arguments leading to \eqref{eq:lambdau_GL} do not hold.
With a detailed DNS study of the momentum- and heat-budgets and the friction factor, and using the wind model of the accompanying paper \cite{vanReeuwijk2007d}, we derive new parameterisations for $\lambda_u$ and $C_f$.

A related question is whether or not the boundary layers can be regarded as turbulent.
The Reynolds number $\Re$ is too low to sustain a "classical" turbulent boundary layer ($\Re \approx 1500$ at $\Ra=10^8$), i.e. a boundary layer where the turbulence production due to shear is in local equilibrium with dissipation.
Hence, the general view is that the boundary layers are laminar, but time-dependent. 
Although time-dependence due to plume impingement and detachment prevents laminarity in the strict
sense, the assumption could be justified if the plumes are passive with respect to the scaling of integral boundary layer parameters such as the friction factor $C_f$ and the kinematic boundary layer thickness $\lambda_u$.
Several other studies show that the friction factor scales similar to a Blasius boundary layer \citep{Chavanne1997, Chavanne2001, Amati2005}.
However, the scaling of $\lambda_u$ does not comply with classical laminar scaling \eqref{eq:lambdau_GL}, as discussed before.
Furthermore, a recent study of time-spectra in the bottom kinematic boundary layer revealed that the spectra in the boundary layer and in the bulk were practically indistinguishable \cite{Verdoold2008}, a strong indication of turbulence.
In order to understand this dual behavior, we study several turbulence indicators for the boundary layers, such as the spectra, shape- and friction factor.

The paper is outlined as follows.
A brief summary of the code for direct simulation and symmetry-accounted ensemble-averaging is given in section \ref{par:methodology}.
The scaling of the boundary layer thickness, the velocity profile, the friction factor and the shape factor are studied in sections \ref{par:bl:int_thickness}, \ref{par:bl:int_velprof} and \ref{par:bl:int_Cf} respectively. 
Then, we study the space and time spectra (section \ref{par:bl:spectra}). In section \ref{par:bl:mombud}, the mean momentum and temperature budgets in the boundary layers are studied to clarify the importance of fluctuations in the boundary layers.
Using the results from the momentum budgets, the friction factor $C_f$ is decomposed in a pressure and momentum-flux contribution in section \ref{par:bl:Cf}.
This leads to the insight that the main contribution is by the pressure gradient.
Using the conceptual wind model derived in the accompanying paper \cite{vanReeuwijk2007d}, scaling laws for $C_f$ and $\lambda_u$ are derived in section \ref{par:model}. As the results show that the flow has many typical features of turbulence but also of laminarity, the interpretation of the results is discussed in section \ref{par:bl:turbulent}. Conclusions are drawn in section \ref{par:conclusions}.

\section{\label{par:methodology}Simulations}

Direct simulation of Rayleigh-B\'{e}nard convection has been performed at $\Ra=\{10^5, 10^6, 10^7, 10^8\}$ and $\Pr=1$ in a $\Gamma=4$ aspect-ratio domain. 
The code is based on a second-order variance-preserving finite-difference discretisation of the three-dimensional Navier-Stokes equations and is fully parallellised. For all simulations, a grid was used sufficient resolution to resolve the smallest turbulent scales, i.e.\ the Kolmogorov scale $\eta_K = (\nu^3/\varepsilon)^{1/4}$ and Corrsin scale $\eta_K = \Pr^{-1/2} \eta_K$.
The top and bottom wall are rigid (no-slip) and of fixed temperature. At the side domain boundaries, periodic boundary conditions are applied. 
For each $\Ra$ except the highest, 400 independent realisations were obtained by performing 10 independent simulations and sampling the velocity and temperature field roughly twice every convective turnover time.
Because of a formidable computational requirements for $\Ra=10^8$, we use this simulation only for the results of Fig.\ \ref{fig:int_bl} and confine the wind-decomposed analysis to the lower $\Ra$ cases, though without loss of generality.

Similar to domains confined by sidewalls, a wind structure
develops also in domains with lateral periodic boundary conditions. 
However, here the wind structure can be located anywhere in the domain since it is not kept in place by sidewalls. 
To extract the wind, symmetry-accounted ensemble-averaging is used \citep{vanReeuwijk2005}, which aligns the wind structure in different realizations before averaging. 
In this way a wind structure can be identified unambiguously for these domains, by which a decomposition in wind and fluctuations becomes possible.
The resulting average velocity and temperature (three-dimensional fields) are denoted respectively by $\sav{u}_i$ and $\sav{\Theta}$. 
The tildes are used to distinguish the conditional average from the standard (long-time, ensemble or plane) average $\av{X}$ which is a function of $z$ only. The symmetry-accounted average can be interpreted exactly as classical Reynolds-averaged results.
For further details we refer to the accompanying paper \citep{vanReeuwijk2007d}. 
\section{Results}

\subsection{\label{par:bl:int_thickness}Boundary layer thickness}

\begin{figure}
  \centering
  \includegraphics{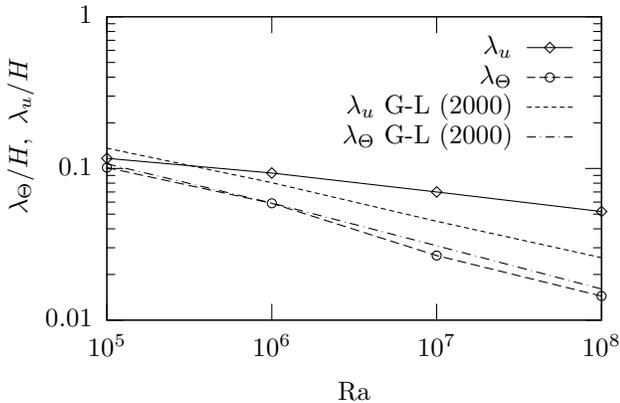}
  \caption{\label{fig:int_bl}
The thermal and kinematic boundary layer thickness $\lambda_\Theta$ and
$\lambda_u$ as a function of $\Ra$.
The dashed and dash-dotted lines in the graphs are predictions by the G-L theory
of $\lambda_u$ and $\lambda_\Theta$, respectively.}
\end{figure}

The thickness of the hydrodynamic and thermal boundary layers as a
function of $\Ra$ is shown in Fig.\ \ref{fig:int_bl}. Here,
$\lambda_u$ and $\lambda_\Theta$ are defined as the location of
the maximum of mean squared horizontal velocity fluctuations
$\av{\fl{u} \fl{u}}$ and mean squared temperature fluctuations
$\av{\fl{\Theta} \fl{\Theta}}$, respectively. The approximate
powerlaws are $\lambda_u = 0.5\ \Ra^{-0.13}$ and $\lambda_\Theta =
2.33\ \Ra^{-0.27}$ respectively, in good agreement with other
simulations \citep{Kerr1996} and reasonable agreement with
experiments \citep{Xin1996} (despite differences in aspect-ratio,
geometry and boundary conditions).

Also shown in Fig.\ \ref{fig:int_bl} are the predictions of the
boundary layer thickness \eqref{eq:lambdat_GL}, \eqref{eq:lambdau_GL} from the Grossmann-Lohse theory \citep{Grossmann2000}, together with the DNS results. 
The thermal boundary layer thickness $\lambda_\Theta$ is in good agreement with the simulations. 
The width of the kinematic boundary layer $\lambda_u$ does not agree so
well with the G-L theory, as the actual exponent is $-0.13$
instead of $-0.25$ (where we have assumed free-fall scaling $\Re \propto \Ra^{1/2}$ for simplicity).

Below we briefly recapitulate the arguments of \cite{Grossmann2004} leading to \eqref{eq:lambdau_GL}.
The starting point is the laminar two-dimensional Prandtl boundary layer equation \cite{White1991, Schlichting2000}
\begin{equation}
  \label{eq:bleqn_lam}
  u \ddx{x}{u} + w \ddx{z}{u} = \nu \ddxsq{z}{u}.
\end{equation}

Upon substituting $x \rightarrow H x$, $z \rightarrow H \Re^{-1/2} z$, $u
\rightarrow U u$ and $w \rightarrow U \Re^{-1/2}  w$, the equations become
parameter independent as
\begin{equation}
  u \ddx{x}{u} + w \ddx{z}{u} = \ddxsq{z}{u}.
\end{equation}
Neither this expression, nor the incompressibility conditions, nor the boundary conditions have an explicit dependence on $\Re$, so the solution has to be independent of $\Re$ as well.
Therefore, the flow pattern undergoes a similarity transformation, and the
boundary layer thickness scales as $\lambda_u / H \propto \Re^{-1/2}$.
This result is rigorous, provided that \eqref{eq:bleqn_lam} holds, i.e.\ that turbulent stresses do not play a role in the momentum budget.
In section \ref{par:bl:mombud} we show that forcing due to Reynolds-stresses cannot be neglected for the boundary layer equations so that the laminarity assumption does not hold.

\begin{figure}
  \centering
  \subfigure[]{\includegraphics[width=80mm]{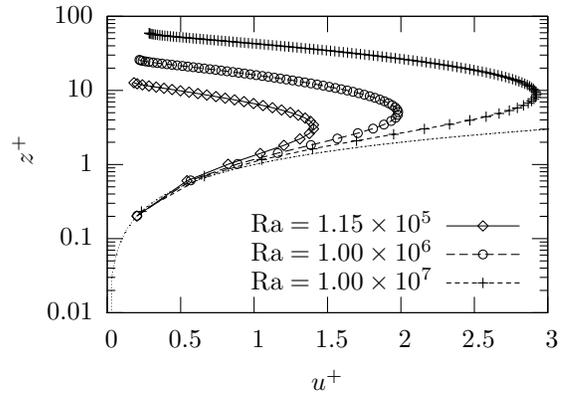}}
  \subfigure[]{\includegraphics[width=80mm]{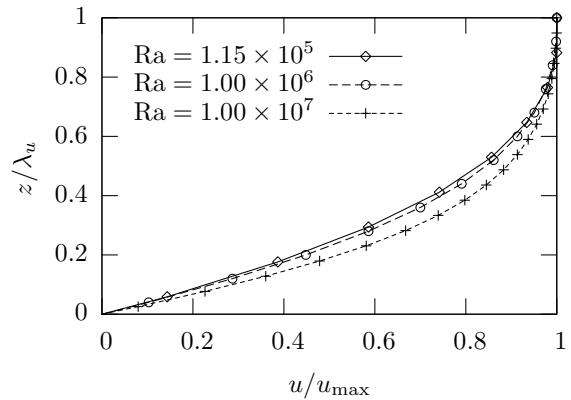}}
\caption{\label{fig:unorm_Ra}
The horizontal boundary layer velocity profile based on $\sqrt{\plav{\sav{u}
~\sav{u}}}$ for various $\Ra$.
a) semi-log plot and non-dimensionalized with friction velocity $u_\tau$.
Dashed line represents $u^+=z^+$;
b) normalized by the maximum velocity $u_{\max}$ and the kinematic boundary
layer thickness $\lambda_u$.
}
\end{figure}

\subsection{\label{par:bl:int_velprof}Velocity profiles}

The characteristic shape of the velocity profile can be obtained
from the plane-averaged horizontal average velocity as $u(z)
\approx \sqrt{\plav{\sav{u} \sav{u}}}$. Figure \ref{fig:unorm_Ra}a
shows these profiles for various $\Ra$ in plus units, i.e.\ scaled
by the friction velocity $u_\tau \equiv \sqrt{\tau_w / \rho}$ with
$u^+=u/u_\tau$ and $z^+ = z u_\tau/\nu$. 
Here, we define a typical wall shear-stress $\tau_w$ as
\begin{equation}
  \tau_w = \rho \nu \ddx{z}{\sqrt{\plav{\sav{u}\sav{u}}}}|_w.
\end{equation}
In Fig.\ \ref{fig:unorm_Ra}a, the viscous sublayer relation
$u^+=z^+$ is shown with a dashed line, and logarithmic scaling of
the velocity profile results in a straight line. For a classical
turbulent channel flow and constant-pressure boundary layer, the
viscous region ends at $y^+=5$, the log-layer starts from
$y^+\approx 30$ and the profiles will collapse onto a single
universal curve for all $\Re$. Here the situation is completely
different. First, in plus-coordinates the profiles do not collapse
onto a single curve. Furthermore, the viscous region ends at
approximately $z^+ = 1$, and the velocity reaches its maximum at
$z^+ \approx 10$ at $\Ra=10^7$. A region where the velocity scales
logarithmically is hard to distinguish, indicating the absence of
an inner (constant stress) layer.

Shown in Fig. \ref{fig:unorm_Ra}b is the velocity profile normalized by the outer variables, i.e.\ the boundary layer thickness $\lambda_u$ and the maximum velocity $u_{\max} \equiv u(\lambda_u)$.
Although the profiles show that there is a $\Ra$ dependence, it is very weak. The weak influence of the $\Ra$ number - especially for the two lower $\Ra$ numbers considered - is further evidence that the kinematic boundary layer does not behave as a classical forced turbulent boundary layer.
Note that the approximate universality of the velocity profiles means that inner and outer variables can be interchanged, in the sense that $\ddx{z}{u}|_w \propto
u_{\max}/\lambda_u$.

Several experiments have shown universality in $\Ra$ upon an outer scaling by boundary layer thickness and maximum velocity \citep{Xin1996, Lam2002, Qiu1998a}, so it is quite interesting that the boundary layer profile found here (Fig.\ \ref{fig:unorm_Ra}b) has a (small) $\Ra$ dependence. There may be several reasons for this difference. The experiments have been carried out at much higher $\Ra$, in the range $\Ra=2\times 10^8 - 9 \times 10^9$ and at higher $\Pr$ (the working fluid was water). 
Furthermore, the presence of side walls and the smaller aspect ratio will be of influence. 
\ignore{An interesting difference between the experiments and our simulations is that the turbulence intensities (relative to the  wind) in the experiment are about 20\%, whereas here they are about 70-80\%.}

It is useful to express the shear-Reynolds number in terms of $\Re$,
$\lambda_u$ and the non-dimensional velocity gradient at the wall.
Let the outer scaled variables be denoted by $\hat{z} \equiv z / \lambda_u$
and $\hat{u} \equiv u / u_{\max}$.
The non-dimensional velocity gradient at the wall is connected to the wall-shear
stress by $\tau_w = \rho \nu u_{\max} \lambda_u^{-1} \ \ddx{\hat{z}}{\hat{u}
|_w}$, where $\ddx{\hat{z}}{\hat{u} |_w}$ is the non-dimensional velocity gradient at the wall.
Hence, the shear Reynolds number can be expressed as
\begin{equation}
  \label{eq:Retau}
  \Re_\tau = \Re^{1/2} \left( \frac{\lambda_u}{H} \right)^{-1/2}
            \left(\ddx{\hat{z}}{\hat{u}|_w} \right)^{1/2}.
\end{equation}
All three terms $\Re$, $\lambda_u$ and $\ddx{\hat{z}}{\hat{u}} |_w$
depend on $\Ra$, although the $\Ra$ dependence of the last term is
very weak as $\ddx{\hat{z}}{\hat{u}|_w} \propto \Ra^{0.06}$.

\renewcommand\arraystretch{2}
\renewcommand\tabcolsep{4mm}
\begin{table}
\centering
\caption{
\label{tab:boundarylayer}
Characteristic numbers for the boundary layer profile
at various $\Ra$: the shear-Reynolds number $\Re_\tau$, the shape factor $S$ and
the friction coefficient $C_f$.}
\begin{tabular}{cccc}
$\Ra$ & $\Re_\tau$ & $C_f$ & $S$ \\ 
\hline
$1.15 \times 10^5$ & 26  & 1.02 & 2.37 \\ 
$1.00 \times 10^6$ & 52  & 0.51 & 2.35 \\ 
$1.00 \times 10^7$ & 119 & 0.23 & 2.27 \\ 
\end{tabular}

\end{table}
\renewcommand\arraystretch{1}

\subsection{\label{par:bl:int_Cf}Friction and shape factor}

The friction and shape factor \citep[e.g.][]{Schlichting2000, White1991} have been calculated for all three $\Ra$ (Table \ref{tab:boundarylayer}).
The friction factor is defined as
\begin{equation}
  \label{eq:Cf}
  C_f = \frac{\tau_w}{\frac{1}{2} u_{\max}^2} = 2 \frac{\Re_\tau^2}{\Re^2}.
\end{equation}
Here we note that combining \eqref{eq:Retau}, \eqref{eq:Cf} and neglecting the small $\Ra$ dependence of the wall-gradient $\ddx{\hat{z}}{\hat{u}} |_w$ gives that $C_f$ can be approximated by
\begin{equation}
  \label{eq:Cf_lambdau}
  C_f \approx \frac{2}{\Re} \left( \frac{\lambda_u}{H} \right)^{-1}.
\end{equation}
This is consistent with the approximation $\tau_w \approx \nu U /
\lambda_u$, which is an important assumption in the
Grossmann-Lohse theory \cite{Grossmann2000}. The observation
\eqref{eq:Cf_lambdau} will prove to be important to establish the
scaling of $\lambda_u / H$ in section \ref{par:model}.

Based on the values of Table \ref{tab:boundarylayer} and in terms
of $\Re$, the friction factor $C_f$ scales as $C_f \propto
\Re^{-0.60}$. An empirical relation for turbulent plane channel
flow is $C_f = 0.073 \Re^{-0.25}$, with $\Re$ based on channel half
width and mean velocity across the channel \citep{Dean1978}. 
The friction factor of laminar boundary layers have a stronger dependence on $\Re$; for plane Poiseuille flow $C_f=8/\Re$ ($\Re$ based on full channel height) and for the Blasius
flat plate flow $C_f = 0.664 \Re_x^{-1/2}$. Hence, judging from the
scaling of friction factor, the behavior of the boundary layer
would be classified as laminar.
These results are consistent with \cite{Chavanne1997, Chavanne2001, Amati2005}.

\begin{figure}
  \centering
  \includegraphics[width=80mm]{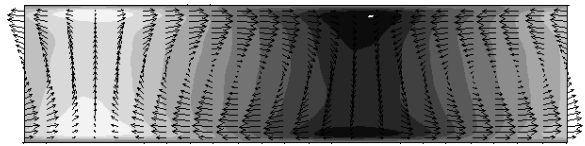}
  \caption{\label{fig:sav}$y$-averaged wind structure at $\Ra=10^6$ and $\Pr=1$. Colorscheme is by the relative temperature $\Theta_r = \sav{\Theta} - \plav{\sav{\Theta}}$. Dark areas are relatively cold, and white areas are relatively hot.}
\end{figure}

The shape-factor $S$ is defined as $S=\delta_1 / \delta_2$, where $\delta_1$
and $\delta_2$ are the displacement and momentum thickness, given
by:
\begin{gather*}
  \delta_1 = \int_0^{\lambda_u} (1-\frac{u}{u_{\max}}) dz, \\
  \delta_2 = \int_0^{\lambda_u} \frac{u}{u_{\max}} (1-\frac{u}{u_{\max}}) dz.
\end{gather*}
For laminar profiles, such as Poiseuille flow and the Blasius
solution for the developing flow over a flat plate, the shape
factor is approximately $2.5$ \citep[e.g.][]{Schlichting2000,
White1991}. For turbulent plane channel flow, flat-plate
constant-pressure boundary layers and a plane turbulent wall jet
\citep{Rajaratnam1976} the shape factor is approximately
$1.3-1.4$. Based on this information, the values from Table
\ref{tab:boundarylayer} indicate that the velocity profile follows
a laminar-like distribution with a slight trend towards turbulent
values as $\Ra$ increases.

If the shape and friction factor are taken to be representative to
distinguish a laminar from a turbulent boundary layer, the
boundary layer would be classified as laminar. In the next
sections we will study the momentum budgets of the boundary
layers, and compare the time and space spectra of boundary layer
and the bulk. It will be shown that from this perspective, the
kinematic boundary layer has many features of turbulence.
\subsection{\label{par:bl:spectra}Fluctuations and spectra}

\begin{figure}
  \centering
    \subfigure[]{\includegraphics[width=52mm]{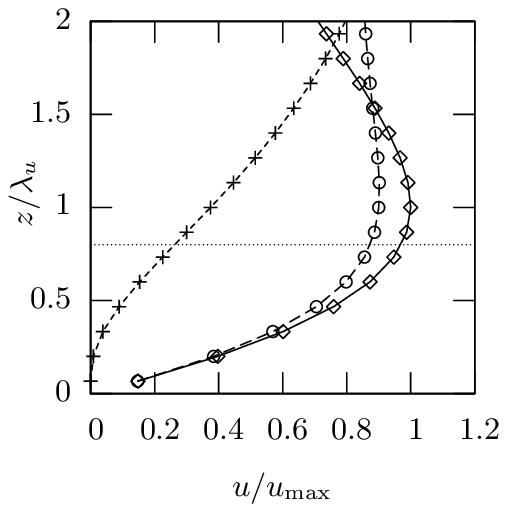}}
    \subfigure[]{\includegraphics[width=52mm]{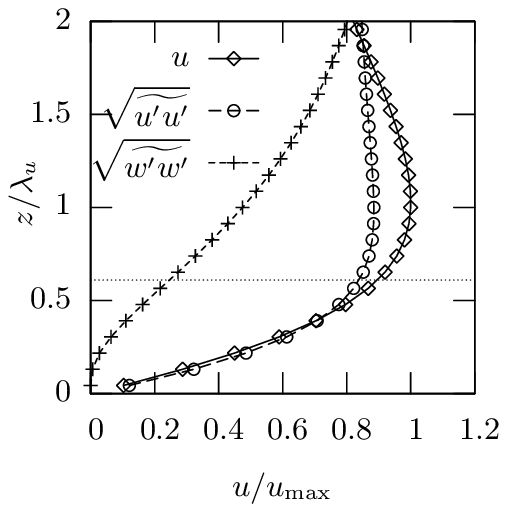}}
    \subfigure[]{\includegraphics[width=52mm]{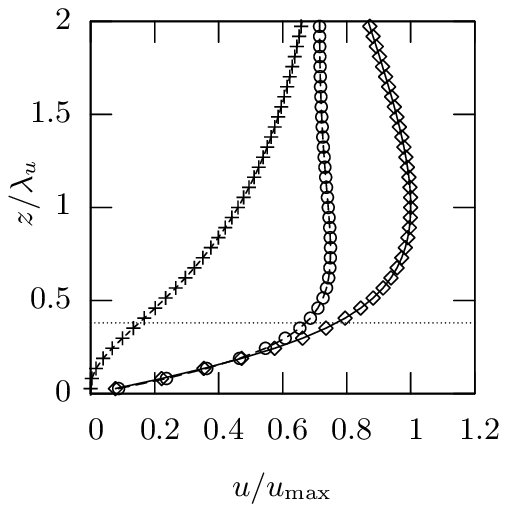}}
  \caption{\label{fig:fluctuations}
Close-up of horizontal velocity and turbulence intensities (legend in Fig.
\ref{fig:fluctuations}b) at the position with the maximum horizontal velocity.
 (a) $\Ra=1.15 \times 10^5$; (b) $\Ra=1.00 \times 10^6$; (c) $\Ra=1.00 \times 10^7$. The horizontal dashed line indicates the edge of the thermal boundary layer $\lambda_\Theta$.}
\end{figure}

In Fig.\ \ref{fig:fluctuations}a-c the average velocity profile $\yav{\sav{u}} / u_{\max}$ is shown for $\Ra=1.15 \times 10^5, 10^6$ and $10^7$, together with the turbulence intensity of the horizontal and vertical fluctuations,
$\yav{\sav{\sfl{u}\sfl{u}}}^{1/2} / u_{\max}$ and
$\yav{\sav{\sfl{w}\sfl{w}}}^{1/2} / u_{\max}$ respectively. These
are the profiles of the $y$-averaged wind structure (see Fig.\
\ref{fig:sav}), with the $x$-location chosen such that the
horizontal velocity is at its maximum, i.e. where the flow is
parallel to the wall and from left to right. A striking feature of
the turbulence intensity of the horizontal fluctuations, is that
it is so large compared to the mean wind, namely 70-80\%. For
turbulent channel flow, typical turbulence intensities are 5-10\%.
Outside the thermal boundary layer the horizontal turbulence
intensity is constant. The vertical turbulence intensity is not as
large as the horizontal due to wall blocking, but is still 20\% at
the edge of the thermal boundary layer, and 50\% at the edge of
the kinematic boundary layer.
This confirms that fluctuations in large aspect-ratio domains are larger relative to the wind \cite{Niemela2006}, in comparison with small aspect-ratio domains (e.g.\ \citep{Xin1996} reports turbulence intensities of 20\%).

One of the main features of turbulence is the presence of a
continuous range of active scales. A simulation at $\Ra=10^7$ is
used to obtain both spatial and temporal spectra of the horizontal
velocity components. To collect temporal spectra, eight points have
been monitored: four bulk and four boundary layer points. The bulk
points are taken at $z_{\mathrm{bulk}}=H/2$ and the boundary layer
points were chosen according to $z_{\mathrm{bl}} = \lambda_u$.
The temporal spectra are generated by segmenting the time series and a
Welch window has been used. Then, averaging was performed over the
spectra of the two horizontal velocity components and the four
monitoring points. The spatial spectra were collected by
performing a 2D FFT and integrating over circles $k_x^2+k_y^2 =
k^2$ and averaging over approximately 10 turnovers.

\begin{figure}
  \centering
  \subfigure[]{\includegraphics[width=75mm]{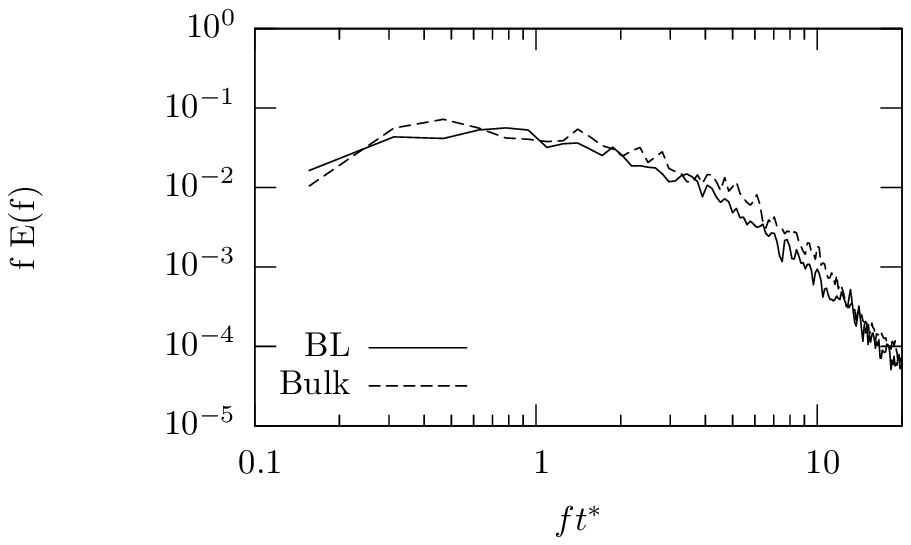}}
  \subfigure[]{\includegraphics[width=75mm]{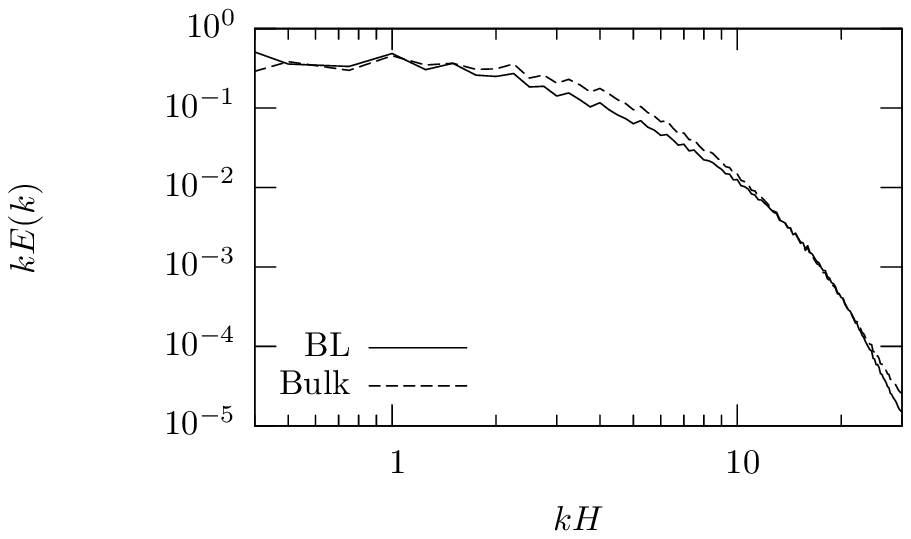}}
  \caption{\label{fig:spectra} Spectra of the horizontal velocity components in the boundary layer and in the bulk at $\Ra=10^7$ normalized by the bulk variance. (a) temporal spectrum (b) spatial spectrum.}
\end{figure}

The temporal spectra of the horizontal velocity components  at
$\Ra=10^7$ are shown in Fig. \ref{fig:spectra}a. There is a
continuous range of active scales which spans about two decades,
although turbulence production and dissipation are not
sufficiently separated to form a clearly discernible inertial
subrange. The spatial spectra (Fig. \ref{fig:spectra}b) also
reveal a continuous range of active scales.

What is striking about the spectra of the bulk and the boundary
layer is how \emph{similar} they are, both in range of active
scales and in amplitude. 
Despite a mild damping at the intermediate frequencies and wave numbers, the similarity indicates that the dynamics of the bulk and the boundary layer - both temporal and spatial - are very similar.
We note that the simulation at $\Ra=10^7$ is well inside the hard-turbulence regime. The transition to hard turbulence occurs at much lower $\Ra$ for large aspect-ratio domains than the generally accepted value of $\Ra = 4 \times 10^7$
\citep{Heslot1987}.
Indeed, for aspect-ratio 6 domains, the flow has hard-turbulence regime scaling occurs from $\Ra = 5 \times 10^4$ upwards \citep{Kerr1996}. 
If one accepts that the flow core is turbulent, then Fig.\ \ref{fig:spectra} indicates that the boundary layers are turbulent as well.

The striking similarity between the spectra in the bulk and the
boundary layers seems to be a robust and general feature of
Rayleigh-B\'{e}nard convection. 
In a recent paper \citep{Verdoold2008}, we present combined experimental and
numerical results of an aspect-ratio 4 cavity filled with water for Rayleigh numbers ranging from $5 \times 10^4$ to $10^9$. 
For all $\Ra$ from $10^6$ upwards, it is found that the spectra in the bulk and the boundary layer are practically identical.
\subsection{\label{par:bl:mombud}Momentum budgets}

\begin{figure*}
  \centering
    \subfigure[$\sav{u}$ budget -$\Ra=10^5$] {\includegraphics[width=55mm]{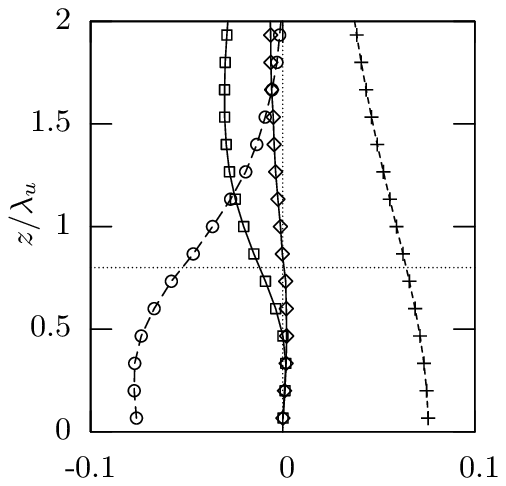}}
    \subfigure[$\sav{u}$ budget -$\Ra=10^6$] {\includegraphics[width=55mm]{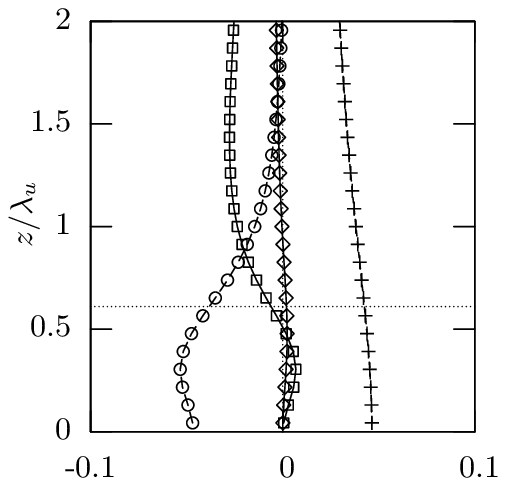}}
    \subfigure[$\sav{u}$ budget -$\Ra=10^7$] {\includegraphics[width=55mm]{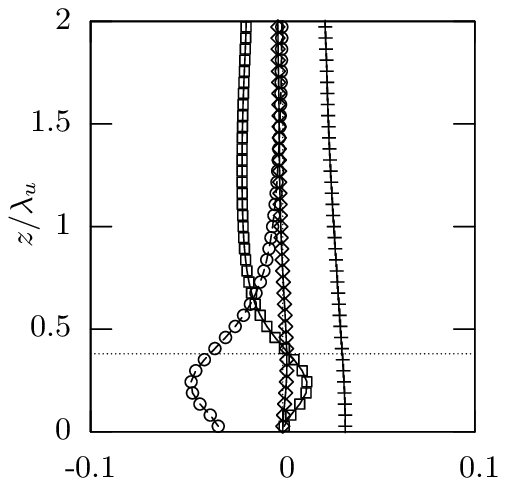}}

    \subfigure[$\sav{w}$ budget -$\Ra=10^5$] {\includegraphics[width=55mm]{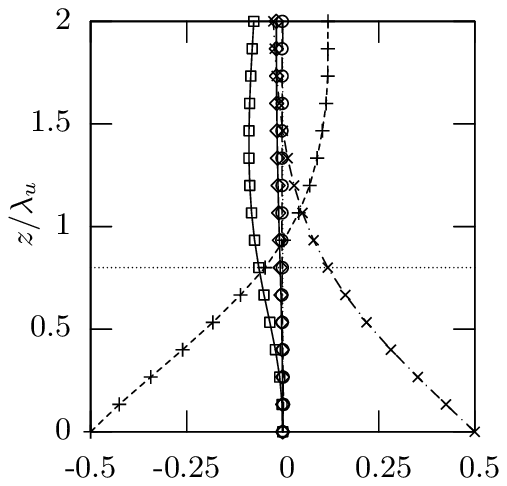}}
    \subfigure[$\sav{w}$ budget -$\Ra=10^6$] {\includegraphics[width=55mm]{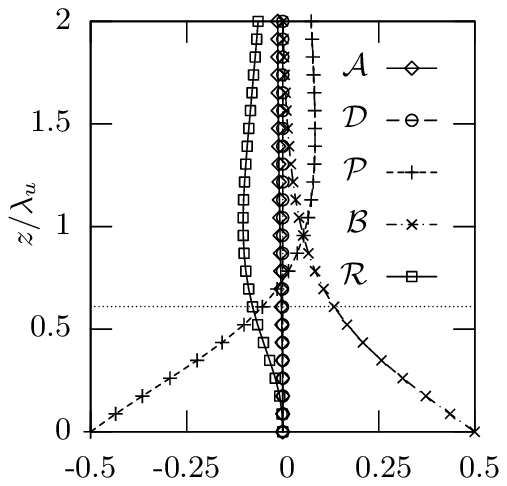}}
    \subfigure[$\sav{w}$ budget -$\Ra=10^7$] {\includegraphics[width=55mm]{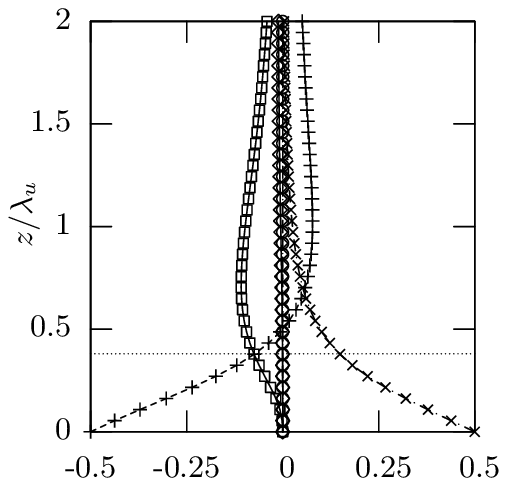}}

    \subfigure[$\sav{\Theta}$ budget -$\Ra=10^5$] {\includegraphics[width=55mm]{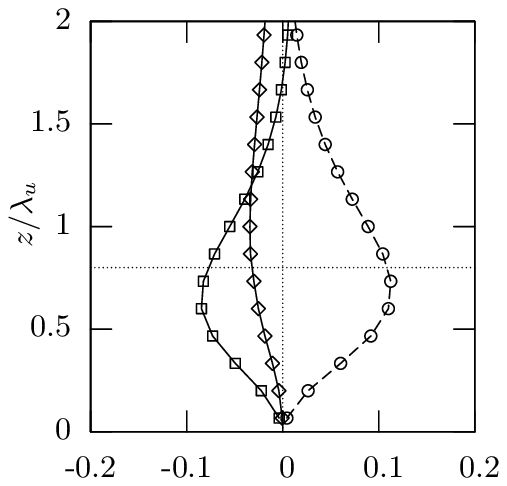}}
    \subfigure[$\sav{\Theta}$ budget -$\Ra=10^6$] {\includegraphics[width=55mm]{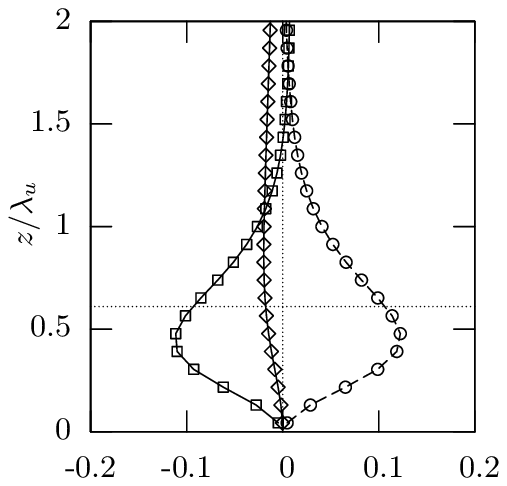}}
    \subfigure[$\sav{\Theta}$ budget -$\Ra=10^7$] {\includegraphics[width=55mm]{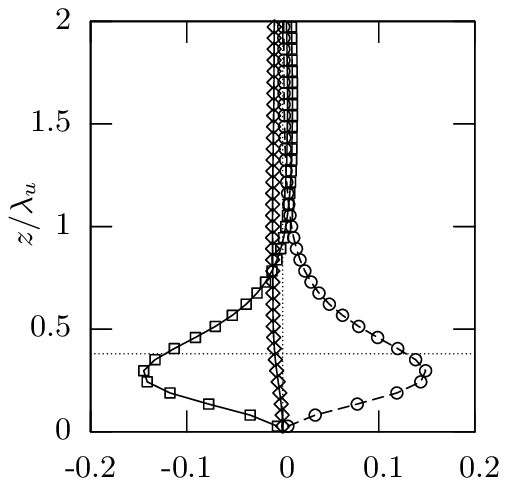}}
  \caption{\label{fig:blprofs}
  Momentum- and heat-budget in the boundary layer at the position with the maximum horizontal velocity for various $\Ra$: (a,d,g): $\Ra=1.15 \times
10^5$; (b,e,h): $\Ra=1.00 \times 10^6$; (c,f,i): $\Ra=1.00 \times 10^7$..
(a,b,c); $\sav{u}$-momentum budget;
(d,e,f): $\sav{w}$-momentum budget;
(g,h,i): $\sav{\Theta}$-budget.
The legend for Figs. (a-i) is shown in Fig. \ref{fig:blprofs}e and
the horizontal dashed line indicates the edge of the thermal boundary layer $z=\lambda_\Theta$.
}
\end{figure*}

Momentum budgets are a very direct way to get an impression of the
importance of the turbulent Reynolds stresses. As before,
$y$-averaged results (Fig. \ref{fig:sav}) are used for convenience
of presentation. Checks have been made to ensure that the budgets
shown here are also representative for the three-dimensional
field. The $x$-location has been chosen such that the horizontal
velocity is at its maximum, i.e. where the flow is parallel to the
wall and from left to right. This guarantees that horizontal
gradients are small, and that no adverse of favorable pressure
gradients are present. Shown are budgets for $\Ra=1.15 \times 10^5$
(Fig \ref{fig:blprofs}a,d,g), $\Ra=1.00 \times 10^6$ (Fig
\ref{fig:blprofs}b,e,h) and $\Ra=1.00 \times 10^7$ (Fig
\ref{fig:blprofs}c,f,i). The budgets for the horizontal (Figs.\
\ref{fig:blprofs}a-c) and vertical momentum (Figs.\
\ref{fig:blprofs}d-f) have been nondimensionalized by $U^2/H=\beta
g \Delta \Theta$, and heat budget (\ref{fig:blprofs}g-i) by
$\Delta \Theta U / H = \sqrt{\beta g (\Delta \Theta)^3/H}$. The
legend for the budgets is shown in Fig. \ref{fig:blprofs}e and the
budget terms are defined in Table \ref{tab:bdg_mom}. The
$z$-coordinate has been scaled by $\lambda_u$ and the horizontal dashed line denotes $z=\lambda_\Theta$.
For reference, the ratio $\lambda_\Theta / \lambda_u$ is 0.8, 0.6 and 0.38 for the simulations at $\Ra=10^5$, $10^6$ and $10^7$ respectively.

\renewcommand\arraystretch{2}
\renewcommand\tabcolsep{1mm}
\begin{table}
 \begin{center}
\caption{\label{tab:bdg_mom} Budget terms for momentum and heat equation.}
\begin{tabular}{cccccc}
& $\bA$ & $\bD$ & $\bP$ & $\bB$ & $\bR$ \\
$\ddt{\sav{u}_i} =$ &
    $-\ddxj{\sav{u}_j \sav{u}_i}$ & $+\nu \ddxjsq{\sav{u_i}}$ &
    $-\ddxi{\sav{p}}$ &
    $+\beta g \sav{\Theta} \delta_{i3}$ & $-\ddxj{\sav{\fl{u}_j \fl{u_i}}}$ \\
$\ddt{\sav{\Theta}} =$ &
    $-\ddxj{\sav{u}_j \sav{\Theta}}$ & $+\kappa \ddxjsq{\sav{\Theta}}$ &&&
    $-\ddxj{\sav{\fl{u}_j \fl{\Theta}}}$

\end{tabular}
\end{center}
\end{table}
\renewcommand\arraystretch{1}

For the horizontal momentum budgets (Figs.\ \ref{fig:blprofs}a-c), the balance is between the horizontal pressure gradient $\bP$ and diffusion $\bD$ for $z < \lambda_\Theta$.
Outside the thermal boundary layer, $\bR$ is not negligible; on the contrary, $\bR$ fully balances the pressure gradient $\bP$ near $z=\lambda_u$. 
This indicates that the turbulence outside the thermal boundary layer is key to the boundary layer thickness, as will be outlined in section \ref{par:model}.
As the location of the budgets has been chosen such that all horizontal
derivatives are small, $\bD \approx \nu \ddxsq{z}{\sav{u}}$ and $\bR \approx -\ddx{z}{\sav{\sfl{w}\sfl{u}}}$.

Log-scaling is expected in the inner layer where $\sav{\sfl{w}\sfl{u}}$ is
constant, so that $\bR = -\ddx{z}{\sav{\sfl{w}\sfl{u}}} = 0$. 
For channel flow, $\bR$ is zero at the wall and peaks in the buffer layer, marking the transport of momentum from the outer to the inner layer. 
After the peak, it crosses the zero axis where the log-layer is expected. This behavior of $\bR$ is absent for $\Ra=1.15 \times 10^5$, but as $\Ra$ increases a peak forms inside the thermal boundary layer (Fig.\ \ref{fig:blprofs}a-c).
However, in terms of forcing $\bP$ is always much larger than the small peak $\bR$ for the range of $\Ra$ under consideration, which again confirms that this is not a classical forced turbulent boundary layer.

Figs.\ \ref{fig:blprofs}d-f show the budgets of the $\sav{w}$-momentum equation.
Here the balance is between buoyancy $\bB$, the vertical pressure gradient
$\bP$ and the Reynolds stresses $\bR = -\ddx{z}{\sav{\sfl{w}\sfl{w}}}$.
Very near the wall, roughly in the lower half of the thermal boundary layer,
the buoyancy $\bB$ and pressure $\bP$ are in balance, so the flow is neutrally
buoyant here.
Further away from the wall, at the edge of the thermal boundary layer, the
contribution of $\bR$ is significant, even if it may seem small compared to the
near-wall (hydrostatic) balance of $\bP$ and $\bB$.
In fact, comparing $\bR$ of the vertical momentum equation to the magnitude of terms in the horizontal momentum equation shows that it is of the same magnitude as $-\ddx{x}{\sav{p}}$.
Outside the boundary layer, the pressure gradient $\bP$ is positive and is balanced purely by fluctuations $\bR$.

The $\sav{\Theta}$-momentum budgets (Figs.\ \ref{fig:blprofs}h-i) show a balance
between thermal diffusion $\bD=\kappa \ddxsq{z}{\sav{\Theta}}$, turbulence
$\bR=-\ddx{z}{\sav{\sfl{w}\sfl{\Theta}}}$ and there is a contribution from
advection $\bA$.
Judging from the peak of $\bA$ around $z/\lambda_u=1$, the nonzero
contribution of $\bA$ to the heat budget is probably caused by some spatial
variations in $\sav{\Theta}$ by which $\ddx{x}{\sav{u}\sav{\Theta}} \neq 0$.
The peak of $\bD$ and $\bR$ is always located just inside the thermal boundary
layer, representing  the location where diffusion and fluctuations most
effectively exchange heat.

It is striking that the dominant length scale for the budgets is the thermal boundary layer thickness $\lambda_\Theta$ (which is denoted by the horizontal dashed line in Fig.\ \ref{fig:blprofs}), and not as one may expect, the kinematic boundary layer thickness. 
Perhaps this should not be too much of a surprise, as the thermal boundary layer thickness can be well represented by $\lambda_\Theta=H/ (2 \Nu)$,
and the Nusselt number $\Nu$ represents the efficiency of the convective heat transfer mechanism of the flow, resulting from the non-linear coupling of temperature and velocity under the action of buoyancy. Therefore, $\lambda_\Theta$ is equally important for the heat-budget and for the momentum budgets. 
In fact, $\lambda_\Theta$ is a dominant parameter in the scaling of both $C_f$ and $\lambda_u$, as will be shown in section \ref{par:model}.

The findings of Figs. \ref{fig:blprofs}a-i can be summarized as follows for the
$\sav{u}$, $\sav{w}$ and $\sav{\Theta}$ budget, respectively:
\begin{gather}
  \label{eq:bleqn_u}
  \ddx{x}{\sav{p}} + \ddx{z}{\sav{\sfl{w}\sfl{u}}} = \nu \ddxsq{z}{\sav{u}}, \\
  \label{eq:bleqn_w}
  \ddx{z}{\sav{p}} + \ddx{z}{\sav{\sfl{w}\sfl{w}}} = \beta g \sav{\Theta}, \\
  \label{eq:bleqn_Theta}
  \ddx{z}{\sav{\sfl{w}\sfl{\Theta}}} = \kappa \ddxsq{z}{\sav{\Theta}}.
\end{gather}
These equations represent the boundary layer equations at the
$x$-location where the flow is parallel to the wall and horizontal
derivatives are negligible (roughly halfway between the
impingement and detachment region). Note that even though the
$\sav{w}$-momentum equation is not directly coupled to the other
two equations, the vertical fluctuations $\sav{\sfl{w}\sfl{w}}$
are non-trivially coupled to $\sav{\sfl{w}\sfl{u}}$ and
$\sav{\sfl{w}\sfl{\Theta}}$ as these terms represent to a large
extent the plumes emerging from and impinging on the boundary
layers. The equations above are two-dimensional, but by the
absence of transversal derivatives, it can be expected that these
equations are valid for the three-dimensional case as well, in a
local coordinate system aligned with the flow and at the location
where the flow is parallel to the wall.

The boundary layer equation \eqref{eq:bleqn_u} clearly shows that one cannot neglect the influence of turbulence in the boundary layer dynamics.
Hence, the laminar boundary layer equation \eqref{eq:bleqn_lam}, which lead to the scaling $\lambda_u \propto \Re^{-1/2}$ is not valid: additional information is required about $\sav{\sfl{w} \sfl{u}}$ to estimate $\lambda_u$. 
In section \ref{par:model}, the scaling behavior of $\lambda_u$ will be derived using flow-specific information obtained from the DNS results.
\subsection{\label{par:bl:Cf}The friction factor decomposed}

By using the boundary layer equation \eqref{eq:bleqn_u}, the
dominant contributor to the friction factor can be identified.
Integrating \eqref{eq:bleqn_u} over the kinematic boundary layer and
substituting \eqref{eq:Cf}, the friction factor $C_f$ is composed
of a contribution from pressure and a turbulent momentum flux as
\begin{equation}
  \label{eq:Cf_decomposition}
  \begin{split}
  \frac{C_f}{2} &= \frac{1}{u_{\max}^2} \int_0^{\lambda_u} (\bP + \bR) dz \\
                &= - \frac{1}{u_{\max}^2} \int_0^{\lambda_u} \ddx{x}{p} dz 
                  - \frac{\sav{\sfl{w}\sfl{u}} |_{\lambda_u}}{u_{\max}^2}.
  \end{split}
\end{equation}
The terms on the right hand side of \eqref{eq:Cf_decomposition} have been calculated with the DNS results and are presented in Table \ref{tab:Cf_decomposition} 
\footnote{The values for $C_f$ in Table \ref{tab:Cf_decomposition} are slightly smaller than those in Table \ref{tab:boundarylayer}, as the latter is deduced from the plane-averaged squared mean velocity.}
The decomposition clearly demonstrates that $C_f$ is dominated by the pressure
gradient. 
The turbulent momentum flux $\sav{\sfl{w}\sfl{u}}$ is small but positive, i.e.\ a flux out of the boundary layer. 
Hence, we conclude that the dynamics of the wall-friction is not governed
by turbulence (i.e.\ Reynolds stress) as in a forced turbulent boundary layer.
In the latter, the dynamics is dominated by a large momentum-flux into the boundary layer, while the contribution of the pressure gradient is negligible.

\renewcommand\arraystretch{2}
\begin{table}
\caption{\label{tab:Cf_decomposition}Decomposition of the friction factor $C_f$
according to \eqref{eq:Cf_decomposition}}.
\begin{tabular}{c|ccc}

$~\Ra~$ & $~~C_f=~~$ & $~~ - \frac{2}{u_{\max}^2} \int_0^{\lambda_u} \ddx{x}{p} dz~~$ &
               $~~- \frac{2 \sav{\sfl{w}\sfl{u}} |_{\lambda_u}}{u_{\max}^2}~~$ \\
\hline
\centering
$10^5$ & 0.81 & 0.88 & -0.07 \\
$10^6$ & 0.39 & 0.42 & -0.03 \\
$10^7$ & 0.17 & 0.19 & -0.02
\end{tabular}
\end{table}
\renewcommand\arraystretch{1}

\begin{figure}
  \includegraphics{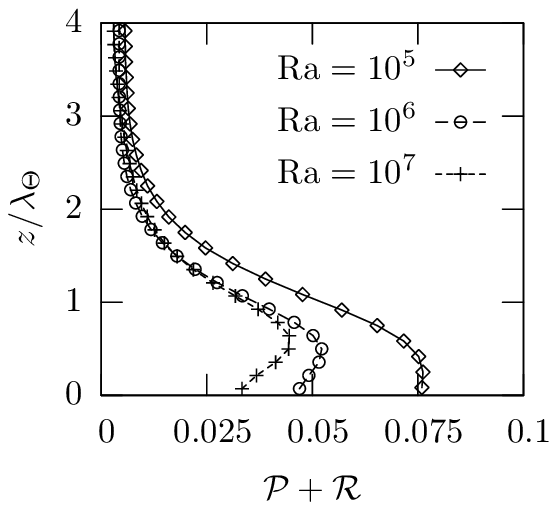}
  \caption{\label{fig:effectiveforce}The effective forcing $\bP+\bR = - \ddx{x}{\sav{p}} - \ddx{z}{\sav{\sfl{w}\sfl{u}}}$, which vanishes quickly outside the thermal boundary layer.}
\end{figure}

Using \eqref{eq:Cf_decomposition}, $C_f$ can be parameterised.
Shown in Fig.\ \ref{fig:effectiveforce} is the effective forcing $\bP + \bR = - \ddx{x}{\sav{p}} - \ddx{z}{\sav{\sfl{w}\sfl{u}}}$.
To first order, for $z<\lambda_\Theta$, $\bP + \bR \approx \bP$ while for $z>\lambda_\Theta$ the Reynolds stress forcing $\bR$ balances the pressure $\bP$ to that $\bP + \bR \approx 0$.
Hence, $C_f$ can be estimated by

\begin{equation}
  \label{eq:Cf_approximation}
  C_f \approx \frac{1}{u_{\max}^2} \int_0^{\lambda_\Theta} \bP dz
      \approx \frac{2 \lambda_\Theta}{u_{\max}^2} \abs{\left. \ddx{x}{p} \right|_w}.
\end{equation}

Clearly, \eqref{eq:Cf_approximation} holds at moderate $\Ra$ only,
when turbulent shear production in the boundary layer is small.
The formation of the peak inside the thermal boundary layer at
$\Ra=10^6$ and $\Ra=10^7$ (Fig.\ \ref{fig:effectiveforce}) suggests
that shear production becomes more important as $\Ra$ increases,
and this will have to be accounted for in \eqref{eq:Cf_approximation} at higher $\Ra$. 
However, it was shown in the accompanying paper \cite{vanReeuwijk2007d} that the wind velocity becomes independent of $C_f$ at sufficiently high $\Ra$, because $C_f$ is negligible compared to the mixing parameter $\alpha$. Therefore, incorrect scaling behavior in \eqref{eq:Cf_approximation} will not influence the wind dynamics at high $\Ra$.
\section{\label{par:model}Scaling of $C_f$ and $\lambda_u$}

Using the simple two-equation wind model derived in the
accompanying paper \cite{vanReeuwijk2007d}, we can establish the
scaling behavior of $C_f$ and $\lambda_u$. The model uses a
dimensionless wind-velocity $\hat{U}_w = U_w / U_f$ and spatial
temperature difference $\hat{\Theta}_w = \Theta_w / \Delta
\Theta$, where $U_f = \sqrt{\beta g \Delta \Theta H}$ is the
free-fall velocity. The governing equations of the model are given
by
\begin{gather}
  \label{eq:windmodel_Uwhat}
  \dddthat{\hat{U}_w} = \frac{2 \hat{L}_w^2}{2 \hat{L}_w^2+1} \left(
     \frac{1}{2 \hat{L}_w} \hat{\Theta}_w
    -(4 \alpha + C_f) \abs{\hat{U}_w} \hat{U}_w \right), \\
  \label{eq:windmodel_Thetawhat}
  \dddthat{\hat{\Theta}_w} =
 \frac{2 \hat{\lambda}_\Theta}{\hat{L}_w} \hat{U}_w
 -\frac{4 \alpha }
       {\hat{L}_w^2 \Pr_T} \abs{\hat{U}_w} \hat{\Theta}_w
 - \frac{2}
               {\hat{\lambda}_\Theta \Re_f \Pr} \hat{\Theta}_w.
\end{gather}
Here $\hat{L}_w = L_w / H$ where $L_w$ is the typical roll size,
$\hat{\lambda}_\Theta = \lambda_\Theta / H$, $\hat{\lambda}_u =
\lambda_u / H$ and $\Re_f = U_f H / \nu = \Ra^{1/2} \Pr^{-1/2}$. The
turbulent Prandtl number $\Pr_T$ and the mixing parameter
$\alpha$ are coefficients with values $0.85$ \cite{Schlichting2000}
and $0.6$ respectively.
The pressure difference which drives the wind is generated by a spatial temperature difference $\Theta_w$ (it is relatively hot where the flow ascends and relatively cold where it descends, see Fig.\ \ref{fig:sav}). 
The temperature difference $\Theta_w$ is in its turn generated by large horizontal heat fluxes originating from the interaction between the mean wind
and temperature field. 
The model depends on $\Ra$, $\Pr$ and $\hat{L}_w$, where $\hat{\lambda}_\Theta =
\hat{\lambda}_\Theta(\Ra, \Pr)$ and $C_f = C_f(\Ra, \Pr)$ have to be provided. 
Based on the analysis of the friction factor (section \ref{par:bl:Cf}), an explicit expression for $C_f$ can be derived, by which the model only depends on empirical input for $\hat{\lambda}_\Theta$ (and thus $\Nu$).

The steady state estimate for the pressure gradient at the bottom
wall of the wind model is \cite{vanReeuwijk2007d}
\begin{equation}
  \label{eq:windmodel_dpdx}
  - \left. \ddx{x}{\sav{p}} \right|_w \approx
  \frac{\beta g H}{L_w} \Theta_w.
\end{equation}
Using \eqref{eq:windmodel_dpdx}, $C_f$ \eqref{eq:Cf_approximation}
can be further specified as
\begin{equation}
  \label{eq:windmodel_Cf}
  C_f \approx \frac{2 \lambda_\Theta}{H} \frac{H}{L_w} \frac{U_f^2}{U_w^2} \frac{\abs{\Theta_w}}{\Delta \Theta}
  = \frac{2 \hat{\lambda}_\Theta \abs{\hat{\Theta}_w}}{\hat{L}_w \hat{U}_w^2}.
\end{equation}
Hence, the wall friction term is linear in the temperature difference as
\begin{equation}
  \label{eq:friction_estimation}
  C_f \abs{\hat{U}_w} \hat{U}_w = \frac{2 \hat{\lambda}_\Theta} {\hat{L}_w} \hat{\Theta}_w.
\end{equation}
Here, we assumed that $\mathrm{sgn}~ \hat{U}_w = \mathrm{sgn}~ \hat{\Theta}_w$. With \eqref{eq:friction_estimation}, the empirical specification of $C_f(\Ra, \Pr)$ is no longer necessary, and the model is given by:
\begin{gather}
  \label{eq:windmodel_Uwhat_noCf}
  \dddthat{\hat{U}_w} = \frac{2 \hat{L}_w^2}{2 \hat{L}_w^2+1} \left(
     \frac{1 - 4 \hat{\lambda}_\Theta}{2 \hat{L}_w} \hat{\Theta}_w
    -4 \alpha \abs{\hat{U}_w} \hat{U}_w \right),
    \\
  \label{eq:windmodel_Thetawhat_noCf}
  \dddthat{\hat{\Theta}_w} =
 \frac{2 \hat{\lambda}_\Theta}{\hat{L}_w} \hat{U}_w
 -\frac{4 \alpha }
       {\hat{L}_w^2 \Pr_T} \abs{\hat{U}_w} \hat{\Theta}_w
 - \frac{2}
               {\hat{\lambda}_\Theta \Re_f \Pr} \hat{\Theta}_w.
\end{gather}

\begin{figure*}
  \subfigure[]{\includegraphics[width=75mm]{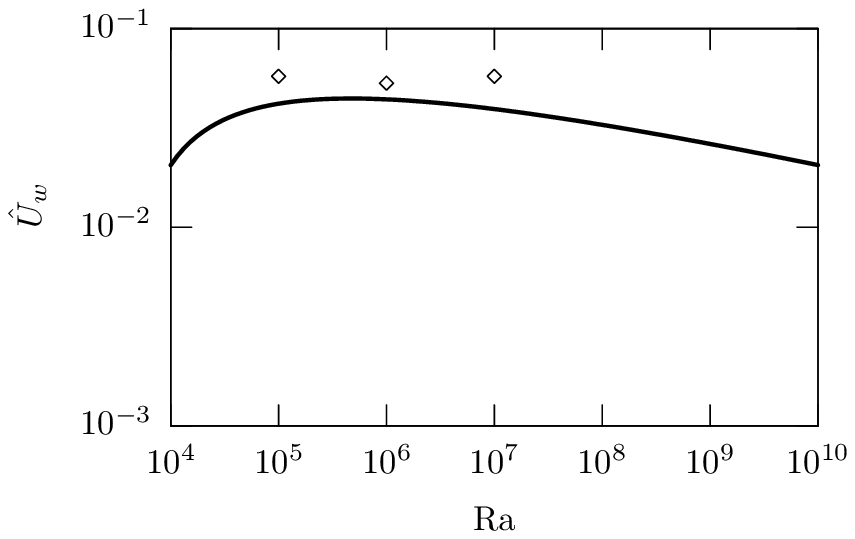}}
  \subfigure[]{\includegraphics[width=75mm]{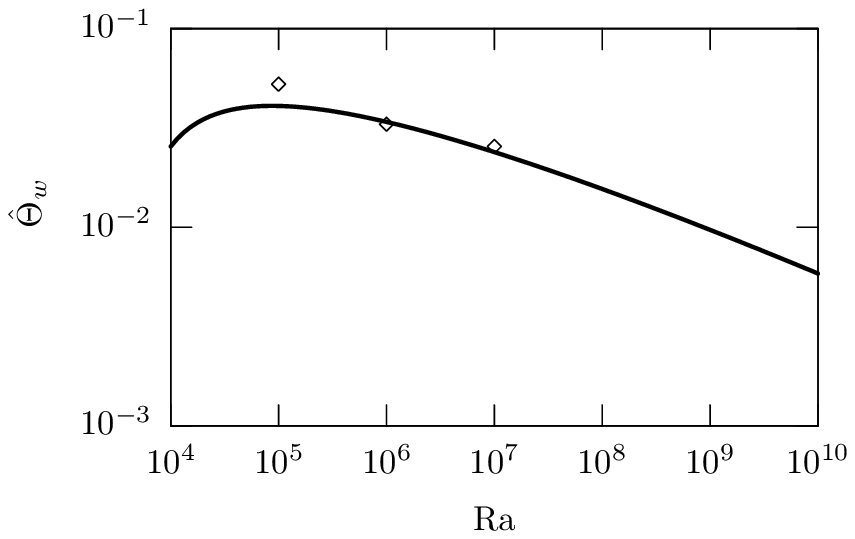}}
  \subfigure[]{\includegraphics[width=75mm]{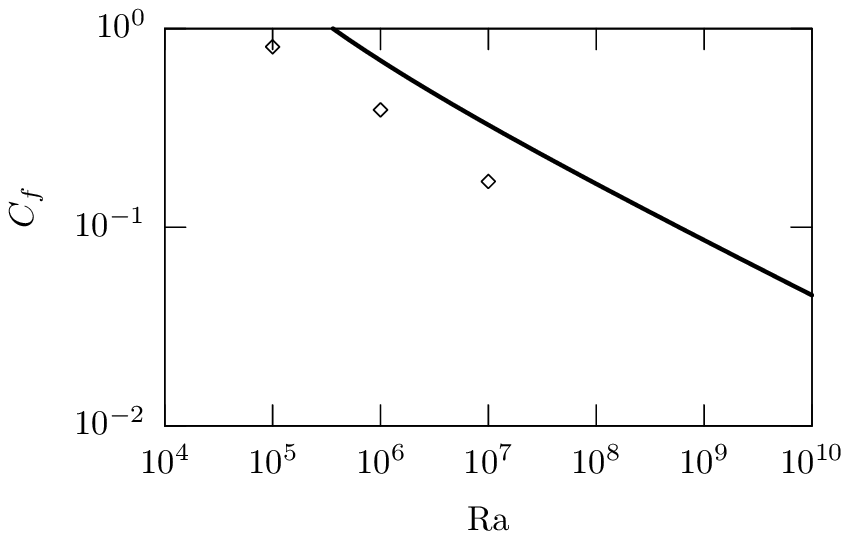}}
  \subfigure[]{\includegraphics[width=75mm]{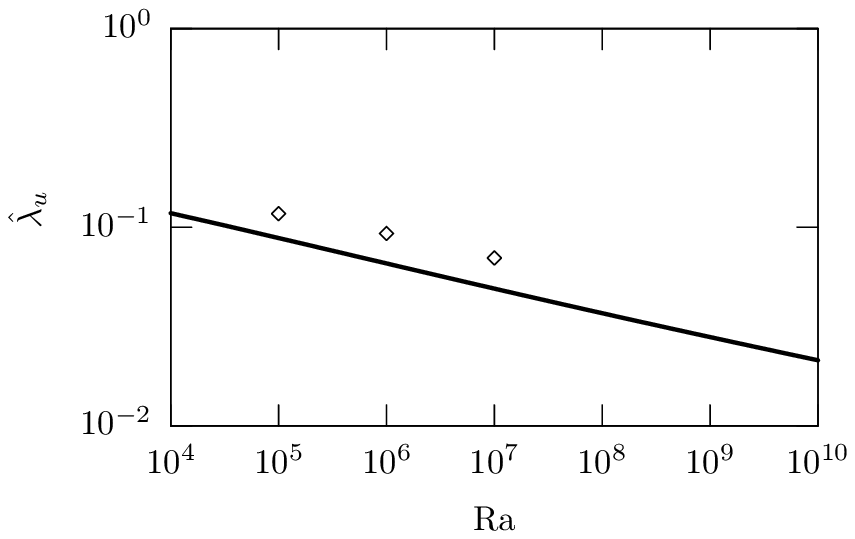}}
\caption{\label{fig:windmodel_noCf}The predictions of the wind model (eqns \eqref{eq:windmodel_Uwhat_noCf}, \eqref{eq:windmodel_Thetawhat_noCf}, thick lines) compared to the DNS results (diamonds) for a) the typical wind $\hat{U}_w$; b) the spatial temperature difference $\hat{\Theta}_w$; c) the friction factor $C_f$ and d) kinematic boundary layer thickness $\lambda_u$.}
\end{figure*}

The steady state solution of the model as a function of $\Ra$ is
shown in Fig.\ \ref{fig:windmodel_noCf}. At this point, the only
empirical data used in the model is the powerlaw for
$\lambda_\Theta$ and the roll size $L_w$. The mixing
parameter $\alpha$ is kept at the same value as in
\cite{vanReeuwijk2007d}, namely 0.6. As can be seen, the model
captures the trends of $\hat{U}_w$, $\hat{\Theta}_w$ and $C_f$
satisfactorily. Note that the profiles could be made to match
quantitatively as well when one would introduce some additional
coefficients. However, the focus of this paper is not to develop a
carefully tuned model, but to elicit general scaling behavior.

It is not very useful to have an expression for $C_f$ in terms of
$\hat{\Theta}_w$, as this quantity is rarely measured. 
However, by using the steady state solution of \eqref{eq:windmodel_Uwhat_noCf}, 
$\hat{\Theta}$ can be expressed in terms of $\hat{U}_w$ as
\begin{equation}
  \label{eq:ThetaUw}
  \hat{\Theta}_w
    = \frac{8 \alpha \hat{L}_w}{1-4 \hat{\lambda}_\Theta} \abs{\hat{U}_w} \hat{U}_w,
\end{equation}
so that the $\hat{\Theta}_w$ dependence of $C_f$ can be eliminated.
Using \eqref{eq:friction_estimation} and \eqref{eq:ThetaUw}, $C_f$ is given by
\begin{equation}
 C_f \approx \frac{16 \alpha }{1-4 \hat{\lambda}_\Theta} \hat{\lambda}_\Theta.
\end{equation}
Hence, when $\hat{\lambda}_\Theta \ll 1$, the model predicts that
$C_f \propto \hat{\lambda}_\Theta$. Note that the $\lambda_\Theta$
term in the denominator represents the effects of wall friction.
Hence, when $\hat{\lambda}_\Theta \ll 1$, $C_f$ scales
\emph{independently} of wall-effects. It is the turbulence in the
outer flow which fully determines the velocity at the edge of the
boundary layer.

A scaling relation for $\lambda_u$ can be derived by using the two
different expressions for $C_f$, \eqref{eq:Cf_lambdau} and
\eqref{eq:windmodel_Cf}. This results in
\begin{equation*}
  2 \Re^{-1} \left( \frac{\lambda_u}{H} \right)^{-1}
  = \frac{2 \hat{\lambda}_\Theta \abs{\hat{\Theta}_w}}{\hat{L}_w \hat{U}_w^2}.
\end{equation*}
Using $\Re = \abs{\hat{U}_w} \Re_f$, $\hat{\lambda}_u$ is approximated by
\begin{equation}
\hat{\lambda}_u \approx \frac{\hat{L}_w \abs{\hat{U}_w}}{\hat{\lambda}_\Theta \abs{\hat{\Theta}_w} \Re_f}
\end{equation}
Dropping the absolute signs and using \eqref{eq:ThetaUw}, $\hat{\lambda}_u$ is given by
\begin{equation}
\hat{\lambda}_u \approx \frac{1-4 \hat{\lambda}_\Theta}{8 \alpha \hat{\lambda}_\Theta \hat{U}_w \Re_f}
         = \frac{1-4 \hat{\lambda}_\Theta}{8 \alpha }
           \frac{1}{\hat{\lambda}_\Theta \Re}.
\end{equation}
Upon assuming that $\hat{\lambda}_\Theta \ll 1$, it follows that
$\hat{\lambda}_u$ scales as  $\hat{\lambda}_u \propto
\hat{\lambda}_\Theta^{-1}  \Re^{-1}$. Fig.\
\ref{fig:windmodel_noCf}d shows the prediction of the wind model
for $\lambda_u$. Although the boundary layer thickness is
underpredicted, the trend is in agreement with the DNS data. Given
the simplicity (with only one calibration parameter $\alpha$), the
model captures the trends of wind velocity, spatial temperature
difference, friction factor and kinematic boundary layer thickness
satisfactorily.
\section{\label{par:bl:turbulent}Turbulent or not?}

The apparently contradicting findings reported in the previous sections is quite intriguing. 
On the one hand, the results indicate that the kinematic boundary layer is turbulent. 
The deduced boundary layer equation \eqref{eq:bleqn_u} shows that forcing due to turbulent Reynolds stresses is significant, in particular outside the thermal boundary layer. Furthermore, the spectra in the bulk and the boundary layers are nearly indistinguishable and show the existence of a continuous range of active scales both in space and time.
Both are an indication for turbulence.

On the other hand, the results suggest that the kinematic boundary layer does not correspond to a classical turbulent boundary layer.
The Reynolds numbers in the $\Ra$ range we consider ($\Re \approx 1500$ at  $\Ra=10^8$)  are generally considered too low to sustain turbulence.
Moreover, the friction factor $C_f$ for a classical forced boundary layer has a weak dependence on $\Re$ (reflecting the quadratic wall friction), and is dominated by the turbulent momentum-flux from the free stream.
For the boundary layers under consideration, the dominant contributor to $C_f$ is the pressure gradient (section \ref{par:bl:Cf}) and not the momentum-flux. Consequently, $C_f$ has a significant $\Re$ dependence. 
The near-universal profiles (found in the present work especially for the two lower $\Ra$ numbers) as a function of $\Ra$ based on the outer variables
$\lambda_u$ and $\sav{u}_{\max}$ (Fig.\ \ref{fig:unorm_Ra}) are further evidence against a classical turbulent boundary layer: 
a turbulent boundary layer can by definition not be universally scaled by outer variables.
\ignore{Although viscous terms are important as can be judged from the low values of $\Re_\tau$ (Table \ref{tab:boundarylayer}), the strong fluctuations resulting from plumes, impinging onto and detaching from the boundary layer, prevent laminarity.}

The difference between classical forced turbulence boundary layers and a boundary layer of Rayleigh-B\'{e}nard convection may be best characterized by the way the turbulence is produced and redistributed.
For forced flow cases, turbulence cannot be maintained at low $\Re$, as the
dissipation in the boundary layer will be stronger than the production.
However, for Rayleigh-B\'{e}nard convection the production and transport of
turbulent kinetic energy (TKE) is not confined to the inner layer alone.
Instead, TKE is produced in the bulk, where it is partially dissipated.
The surplus is transported to the boundary layer by pressure velocity
fluctuations \citep[see also ][]{Kerr2001}.
Therefore, there is no local equilibrium between production and dissipation, and turbulence can be maintained in the boundary layers below the critical $\Re$.
At sufficiently high $\Ra$, instabilities due to shear can be expected to
maintain themselves, and several experiments and simulations show such a
transition around $\Ra=10^{11}$ \citep{Chavanne2001, Niemela2003, Amati2005}.

A simple explanation for the laminar-like scaling of classical integral boundary layer parameters may be that the forcing in the wall-parallel direction is very weak compared to the forcing in the wall-normal direction (plume impingement and detachment).
Indeed, the forcing in the vertical direction is the direct result of buoyancy, while the pressure gradient in the horizontal direction forms is due to large-scale differences in mean temperature.
This can be made explicit by considering the ratio of forcing in the wall-normal direction (buoyancy) and wall-parallel direction \eqref{eq:windmodel_dpdx}, which is given by
\begin{equation}
  \frac{\beta g \Delta \Theta }{\beta g H L_w^{-1} \Theta_w}
= \hat{L}_w \hat{\Theta}_w^{-1}.
\end{equation}
At $\Ra=10^5$, this ratio is approximately 50, and at $\Ra=10^7$,
the ratio is approximately 100. Thus, the boundary layers under
consideration here are forced primarily in the wall-normal
direction, and the force generating the wind is relatively weak.

Despite the laminar-like scaling of the integral parameters, a parallel can be drawn with a fully developed forced boundary layer: both have a viscous sublayer dominated by viscosity which suppresses instabilities and prevents their growth and development of turbulence. 
However, as demonstrated by seminal experiments in the sixties \cite{Kline1967}, despite linear velocity variation, the flow within the sublayer in a forced boundary layer is not laminar, but accompanied by considerable irregular fluctuations, streaks and other structures. 
One can argue that the same dynamics occur in the boundary layers of Rayleigh-B\'{e}nard convection.
In particular, Figs. \ref{fig:blprofs}a-c indicate that the thermal boundary layer $0 < z < \lambda_\Theta$ functions as a viscous sublayer, and the region $\lambda_\Theta < z < \lambda_u$ as an cross-over region between the exterior flow and the thermal boundary layer.
The absence of a constant stress layer dominated by the turbulent momentum flux suppresses a logarithmic region and marks a fundamental difference with forced turbulent boundary layers.

\section{\label{par:conclusions}Conclusions}

The aim of this paper has been to study the boundary layers which develop under the joint action of plumes and wind in Rayleigh-B\'{e}nard convection at the top and bottom plates. 
Direct numerical simulation was used for simulations at $\Ra=\{ 10^5, 10^6, 10^7, 10^8 \}$ and $\Pr=1$ for $\Gamma=4$ aspect ratio domains with periodic side
boundary conditions. For each $\Ra$, 10 independent simulations
have been carried out, resulting in approximately 400 independent
realizations per $\Ra$. Processing the results using
symmetry-accounting ensemble averaging made it possible to retain
the wind structure, which would normally cancel out due to the
translational invariance of the system.

The importance of Reynolds-stresses in the boundary layers, as well as the temporal and spatial spectra indicate undoubtedly a turbulent character of the boundary layer.
However, the behavior is rather different from classical forced boundary layers, as can be judged from the laminar-like scaling of the classical integral boundary layer parameters.
Indeed, viscous effects play an important role within the thermal boundary layer, and a large turbulent momentum-flux from the external stream is absent.
This difference is probably caused by the fact that the turbulence inside the kinematic boundary layer of RB originates from the bulk, whereas classical forced boundary layers are in a local equilibrium between production and dissipation of turbulent kinetic energy.

Due to the importance of Reynolds stresses in the boundary layer, the arguments underpinning the kinematic boundary layer scaling $\lambda_u \propto \Re^{-1/2}$ do not hold.
Using the DNS results and a conceptual wind model \cite{vanReeuwijk2007d}, explicit expressions for $C_f$ and $\lambda_u$ were derived.
It was found that the friction factor should scale proportional to the thermal boundary layer thickness as $C_f \propto \lambda_\Theta$.
The kinematic boundary layer thickness $\lambda_u$ scales inversely proportional to the thermal boundary layer thickness and the Reynolds number as $\lambda_u \propto \Re^{-1} \lambda_\Theta^{-1}$.
The predicted trends for $C_f$ and $\lambda_u$ are in agreement with the DNS results.

With the closure for $C_f$, the model \eqref{eq:windmodel_Uwhat_noCf},
\eqref{eq:windmodel_Thetawhat_noCf} depends solely on empirical input for $\lambda_\Theta$, and predicts the wind Reynolds number $\Re$, friction factor $C_f$ and kinematic boundary layer thickness $\lambda_u$.

\begin{acknowledgments}
This work is part of the research programme of the Stichting voor Fundamenteel
Onderzoek der Materie (FOM), which is financially supported by the Nederlandse
Organisatie voor Wetenschappelijk Onderzoek (NWO).
The computations were sponsored by the Stichting Nationale Computerfaciliteiten
(NCF).
\end{acknowledgments}

\bibliography{bib}
\bibliographystyle{apsrev}

\end{document}